\documentclass[aps,prb,twocolumn,preprintnumbers,amsmath,amssymb,superscriptaddress]{revtex4} %\usepackage{bbding}
\usepackage{graphicx}
\usepackage{bm}
\usepackage[final]{pdfpages}

\begin{document}
\title{Protected gap closing in Josephson junctions constructed on Bi$_2$Te$_3$ surface}
\date{\today} %It is always \today, but any date may be explicitly specified
%\pacs{74.45.+c, 03.65.Vf, 71.70.Ej, 73.40.-c} % PACS, the Physics and Astronomy Classification Scheme.
%\keywords{topological insulators, superconducting proximity effect, unconventional superconductivity}

\author{Zhaozheng Lyu}\thanks{These authors contribute equally to this work.}
\affiliation{Beijing National Laboratory for Condensed Matter Physics, Institute of Physics, Chinese Academy of Sciences; School of Physical Sciences, University of Chinese Academy of Sciences, Beijing 100190, People's Republic of China}
\author{Yuan Pang}\thanks{These authors contribute equally to this work.}
\affiliation{Beijing National Laboratory for Condensed Matter Physics, Institute of Physics, Chinese Academy of Sciences; School of Physical Sciences, University of Chinese Academy of Sciences, Beijing 100190, People's Republic of China}
\author{Junhua Wang}\thanks{These authors contribute equally to this work.}
\affiliation{Beijing National Laboratory for Condensed Matter Physics, Institute of Physics, Chinese Academy of Sciences; School of Physical Sciences, University of Chinese Academy of Sciences, Beijing 100190, People's Republic of China}
\author{Guang Yang}
\affiliation{Beijing National Laboratory for Condensed Matter Physics, Institute of Physics, Chinese Academy of Sciences; School of Physical Sciences, University of Chinese Academy of Sciences, Beijing 100190, People's Republic of China}
\author{Jie Fan}
\affiliation{Beijing National Laboratory for Condensed Matter Physics, Institute of Physics, Chinese Academy of Sciences; School of Physical Sciences, University of Chinese Academy of Sciences, Beijing 100190, People's Republic of China}
\author{Guangtong Liu}
\affiliation{Beijing National Laboratory for Condensed Matter Physics, Institute of Physics, Chinese Academy of Sciences; School of Physical Sciences, University of Chinese Academy of Sciences, Beijing 100190, People's Republic of China}
\author{Zhongqing Ji}
\affiliation{Beijing National Laboratory for Condensed Matter Physics, Institute of Physics, Chinese Academy of Sciences; School of Physical Sciences, University of Chinese Academy of Sciences, Beijing 100190, People's Republic of China}
\author{Xiunian Jing}
\affiliation{Beijing National Laboratory for Condensed Matter Physics, Institute of Physics, Chinese Academy of Sciences; School of Physical Sciences, University of Chinese Academy of Sciences, Beijing 100190, People's Republic of China}\affiliation{Collaborative Innovation Center of Quantum Matter, Beijing 100871, People's Republic of China}
\author{Changli Yang}
\affiliation{Beijing National Laboratory for Condensed Matter Physics, Institute of Physics, Chinese Academy of Sciences; School of Physical Sciences, University of Chinese Academy of Sciences, Beijing 100190, People's Republic of China}\affiliation{Collaborative Innovation Center of Quantum Matter, Beijing 100871, People's Republic of China}
\author{Fanming Qu}
\affiliation{Beijing National Laboratory for Condensed Matter Physics, Institute of Physics, Chinese Academy of Sciences; School of Physical Sciences, University of Chinese Academy of Sciences, Beijing 100190, People's Republic of China}
\affiliation{CAS Center for Excellence in Topological Quantum Computation, University of Chinese Academy of Sciences, Beijing 100190, People's Republic of China}
\author{Li Lu} \email[Corresponding authors: ]{lilu@iphy.ac.cn}
\affiliation{Beijing National Laboratory for Condensed Matter Physics, Institute of Physics, Chinese Academy of Sciences; School of Physical Sciences, University of Chinese Academy of Sciences, Beijing 100190, People's Republic of China}\affiliation{Collaborative Innovation Center of Quantum Matter, Beijing 100871, People's Republic of China}

%\begin{figure}[b]
%\vspace{3 cm}
%\includegraphics[width=1 \linewidth]{cover}
%\end{figure}

\begin{abstract}
On the road of searching for Majorana zero modes (MZMs) in topological insulator-based Josephson junctions, a highly-sought signature is the protected full transparency of electron transport through the junctions due to the existence of the MZMs, associated with complete gap closing between the electron-like and hole-like Andreev bound states (ABSs). Here, we present direct experimental evidence of gap closing and full transparency in single Josephson junctions constructed on the surface of three-dimensional topological insulator (3D TI) Bi$_2$Te$_3$. Our results demonstrate that the 2D surface of 3D TIs provides a promising platform for hosting and manipulating MZMs.
\end{abstract}

\maketitle

\section{INTRODUCTION}

Recently, much attention has been focused on searching for Majorana zero modes (MZMs) in condensed matter systems \cite{Majorana,Majorana return,Search for Majorna,Race for Majorana}. These MZMs, which could be used as topologically protected qubits, are expected to occur at the boundaries of $p$-wave-like superconductors. Experimentally, a number of peculiar phenomena were observed and believed to relate to the occurrence of MZMs, including the appearance of a zero-bias conductance peak (ZBCP) \cite{ZBCP Kouwenhoven,ZBCP H.Q.Xu,ZBCP shtrikman,ZBCP Marcus,ZBCP van Harlingen,ZBCP Yazdani,ZBCP Ando,ZBCP Marcus_2,ZBCP Fan Yang}, the signatures of fractional Josephson effect \cite{Rokhinson,Wiedenmann}, and the emerging of skewed current-phase relations (CPRs) in related Josephson devices \cite{phase-sensitive10 and QP2 Yacoby,Moler_PRL,phase-sensitive7 van Harlingen NatureComm,Pang_CPB}. In the presence of MZMs, however, the ZBCP is expected to be completely quantized, and the charge transport fully transparent (i.e., perfect Andreev reflection) in related devices \cite{4pi1 A Yu Kitaev,4pi2 LowTempPhy,Fu 2008 PRL,4pi3 and rf1 Fu Liang 2009 PRB,4pi4 and rf2 nanowire,4pi5 and rf3 Beenakker,current reverse and QP1,Fu 2013 PRB,4pi6 and rf4 Kane}. While the complete quantization of ZBCP has been observed very recently \cite{ZBCP Hao Zhang}, the complete closing of minigap between the electron-like and hole-like ABSs, as signature of fully transparent charge transport, has not yet been strictly proven.

In this work, we studied the magnetic flux-dependent evolution of the local induced minigap in single Josephson junctions constructed on the surface of three-dimensional topological insulator Bi$_2$Te$_3$. Our results lead to the conclusion that the minigap undergoes complete closing.

\section{EXPERIMENTAL METHOD AND RESULTS}

\begin{figure}
\includegraphics[width=1 \linewidth]{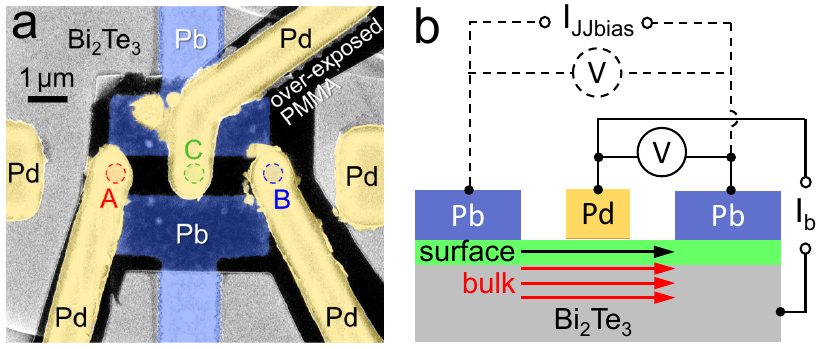}
\caption{\label{fig:fig1} {(color online) (\textbf{a}) The false colored scanning electron microscope image of device \#1. (\textbf{b}) A side view of the device showing that there are two supercurrents flowing through the surface and the bulk of Bi$_2$Te$_3$. Also shown are the wiring configurations for measuring the Fraunhofer pattern (dashed lines) and the contact resistance (solid lines). }}
\end{figure}

The devices used in this work were fabricated on exfoliated Bi$_2$Te$_3$ flakes of $\sim$100 nm in thickness. Two ``T"-shape Pb pads were sputtering deposited on the surface of the flakes to define a superconductor-TI-superconductor (S-TI-S) Josephson junction. Three non-superconducting Pd electrodes were e-beam evaporated and introduced to the Bi$_2$Te$_3$ surface at the left, right and center of the junction (marked by A, B, and C in Fig. 1(a), respectively) for detecting the local electron states. The contacting area of each Pd electrode is defined by the windows through an over-exposed polymethyl methacrylate (PMMA) layer which was fabricated prior the deposition of the Pd electrodes, for the purpose to isolate the arms of the Pd electrodes from the Pb junction underneath. The other two Pd electrodes away from the junction were added for transport measurements. The electron transport measurements with configurations illustrated in Fig. 1(b) were performed by using lock-in amplifier technique, down to 20 mK in a dilution refrigerator. The electron temperature of the system is slightly higher, $\sim$50 mK, known from the numerical simulations for the experimental data (will be shown later).

\subsection{The Josephson supercurrent between the two Pb electrodes}

\begin{figure}
\includegraphics[width=1 \linewidth]{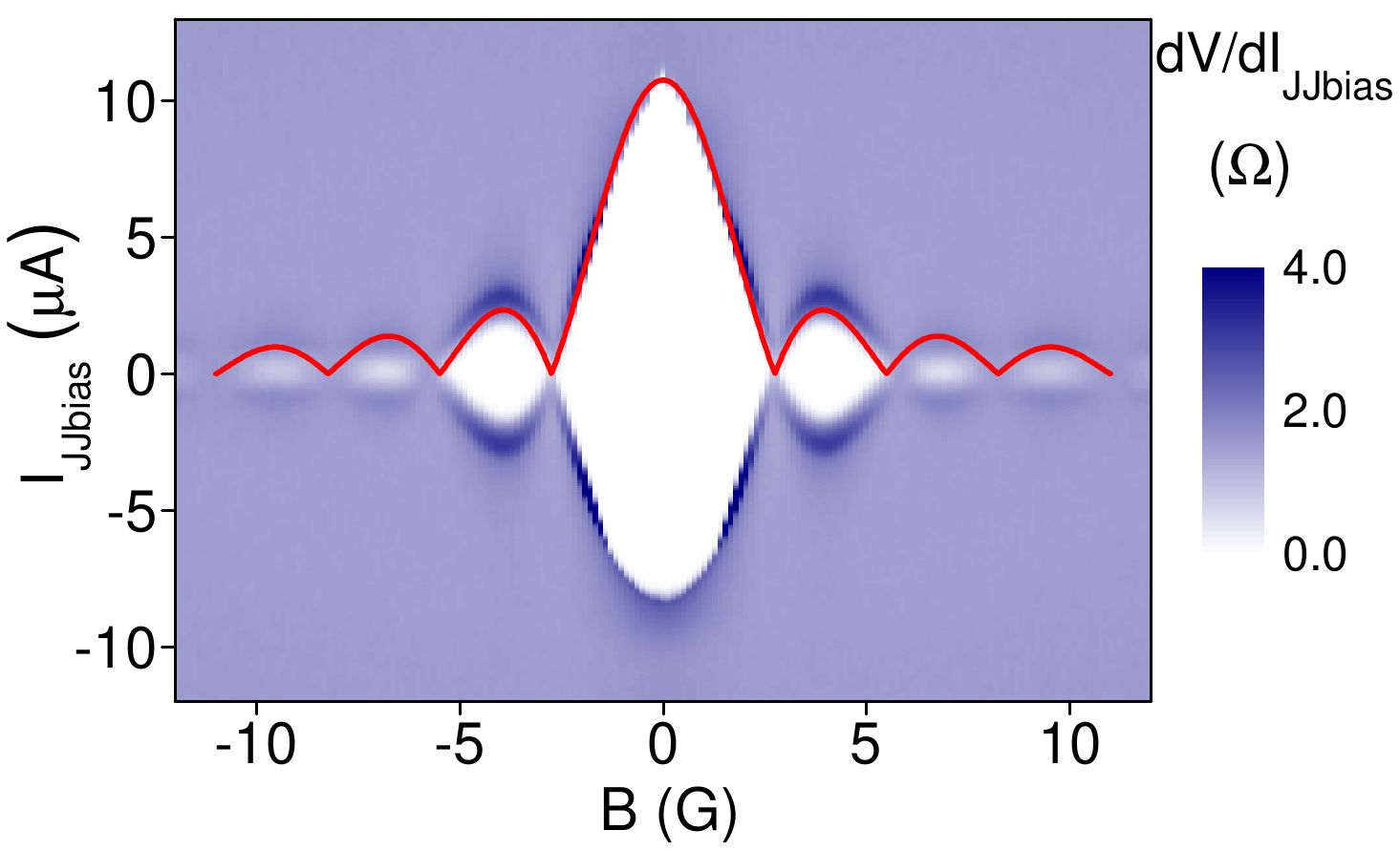}
\caption{\label{fig:fig2} {(color online) The differential resistance $dV/dI_{\rm JJbias}$ measured between the two Pb electrodes, as functions of magnetic field $B$ and bias current $I_{\rm JJbias}$ at $\sim$20 mK. The red line represents the expected Fraunhofer envelope of the zero-resistance state in the non-transparent limit. }}
\end{figure}

\begin{figure*}
\includegraphics[width=1 \linewidth]{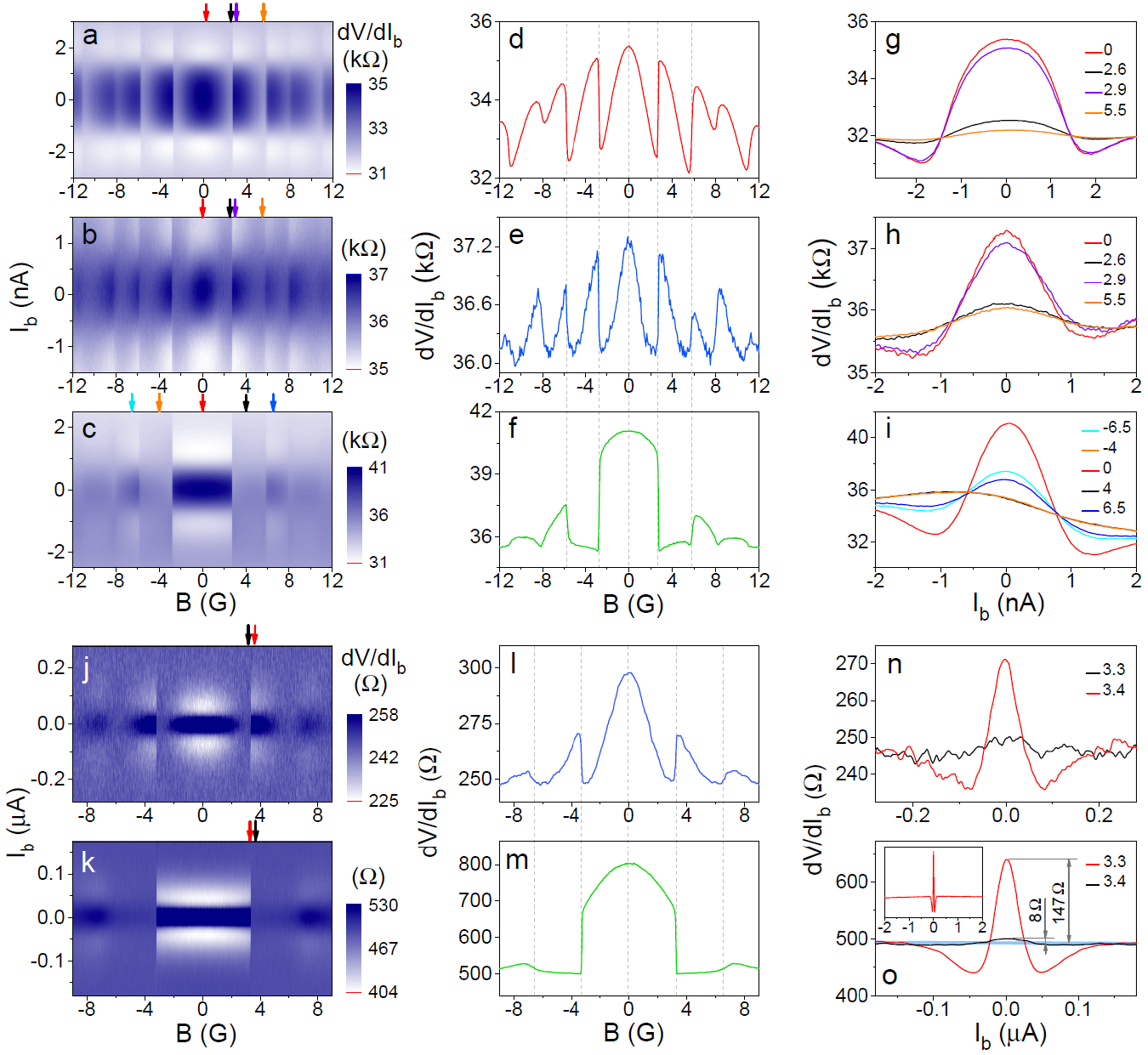}
\caption{\label{fig:fig3} {(color online)
The contact resistance $dV/dI_{\rm b}$ of devices \#1 and \#2 at $\sim$20 mK. (\textbf{a}), (\textbf{b}) and (\textbf{c}) The 2D plots of $dV/dI_{\rm b}$ of device \#1 at positions A, B and C, respectively, as functions of magnetic field $B$ and bias current $I_{\rm b}$. (\textbf{d}), (\textbf{e}) and (\textbf{f}) $dV/dI_{\rm b}$ as a function of $B$ along the horizontal line cuts at $I_{\rm b}=0$ in (a), (b) and (c), respectively. (\textbf{g}), (\textbf{h}) and (\textbf{i}) $dV/dI_{\rm b}$ as a function of $I_{\rm b}$ along the vertical line cuts in (a), (b) and (c), respectively, at magnetic fields indicated by the arrows of corresponding color in the 2D plots. The numbers in the legends are in the unit of Gauss. (\textbf{j})--(\textbf{o}) Similar data obtained at one of the ends and the center of device \#2. The inset in (o) shows the data over a much wider current (hence voltage) range.}}
\end{figure*}

Firstly, we measured the differential resistance $dV/dI_{\rm JJbias}$ between the two Pb electrodes of a Pb-Bi$_2$Te$_3$-Pb Josephson junction (device \#1), as functions of magnetic field $B$ and bias current $I_{\rm JJbias}$ at the base temperature. The results are shown in Fig. 2, with the bias current swept from the bottom to the top. The white-colored area represents the zero-resistance state. The pattern is slightly asymmetric along the vertical direction, as is usually seen in literature for Josephson junctions of the similar type. The envelope of the zero-resistance state along the positive bias direction agrees well with the Fraunhofer pattern of critical supercurrent for Josephson junctions in the non-transparent limit (the red curve, for further explanation please refer to the supplementary materials \cite{supplementary_materials}). The period of the Fraunhofer pattern is $\Delta B=2.8\pm0.2$ G. It corresponds to an effective junction area of $S_{\rm eff}=\phi_0 /\Delta B=7.14 \mu$m$^2$ (where $\phi_0=h/2e$ is flux quantum, $e$ the electron charge, and $h$ the Planck constant). This estimated area is in good agreement with the geometric area of the junction $W\times H=4 \mu m\times 1.8 \mu m=7.2 \mu$m$^2$, where $W$ is the width and $H$ the effective distance between the two Pb electrodes after considering flux penetration and compression \cite{supplementary_materials}.

\subsection{The contact resistance in the junction area probed by Pd electrods}

Then, we measured the contact resistance of the Pd electrodes by employing a three-terminal configuration as illustrated in Fig. 1(b), to probe the local electron states in the junction area. The measurement current was kept $\sim$ 1/1000 of the critical supercurrent of the Josephson junction, to avoid disturbing the status of the Josephson junction. Figures 3(a), (b) and (c) show the 2D color plots of the contact resistance $dV/dI_{\rm b}$ on device \#1 at positions A, B and C, respectively, as functions of $B$ and bias current $I_{\rm b}$. In contrast to the continuous variation of the Fraunhofer pattern shown in Fig. 2, the contact resistance of all contacts exhibits sharp jumps when $B$ is swept across the nodes of the Fraunhofer pattern. These sharp jumps can also be clearly seen from the horizontal line cuts of the 2D plots. The line cuts of Figs. 3(a) and (b) at $I_{\rm b}=0$, plotted in Figs. 3(d) and (e) respectively, show a skewed $B$ dependence followed by abrupt jumps. And the line cut of Fig. 3(c) at $I_{\rm b}=0$, plotted in Fig. 3(f), shows a square-shape $B$ dependence before rectified by the Fraunhofer-like envelope.

In Figs. 3(g), (h) and (i) we plot the vertical line cuts taken at different $B$ in the corresponding 2D plots. It can be seen that a gap-like structure on $dV/dI_{\rm b}-I_{\rm b}$ curves undergoes sudden opening/closing with varying $B$ at the Fraunhofer nodes (i.e., jumping between curves with most pronounced gap structures to least pronounced ones). We will show later that the remaining tiny signature of gap after gap closing, i.e., those on the black and the orange curves in Figs. 3(g), (h) and (i), is caused by the finite size effect of the Pd electrodes.

The contact resistance of Pd electrodes in device \#1 was higher than the quantum unit $h/2e^2=12.9 $k$\Omega$, indicating that the measurement was in the tunneling regime, so that the results reflect mostly the information of the electron density of states underneath. We have fabricated and measured more than ten devices. Similar results were obtained even when the contact resistance of Pd electrodes was in a range below the quantum unit $h/2e^2$, presumably because multiple tunneling channels shunt together in the contacting area. Figures 3(j)--(o) give such an example.

Gap closing can be most clearly seen in the main frame and the inset of Fig. 3(o), where the normal-state value of $dV/dI_{\rm b}$ can be firmly determined and is represented by the horizontal blue line. It can be seen that $\sim$95\% of the peak height suddenly vanishes upon gap closing, i.e., from 147 $\Omega$ at 3.3 G to 8 $\Omega$ at 3.4 G.

Gap closing can also be seen in Figs. 3(d), (e), (f), (i) and (m). The line cuts in these plots, which represent the field-dependent peak height of the $dV/dI_{\rm b}-I_{\rm b}$ curvrs, approach to the normal-state values at their oscillatory minimums.

\section{DISCUSSIONS}
\subsection{The phase dependence of superconducting minigap}

That the contact resistance $dV/dI_{\rm b}$ and the Fraunhofer pattern of supercurrents share the same oscillation period implies that the $dV/dI_{\rm b}$ oscillation is controlled by the same phase difference $\varphi$ across the junction. It is known for S-N-S type Josephson junctions (where N denotes normal metals) that the backwards and forwards Andreev reflections between the two S-N interfaces lead to the formation of ABSs in the N. The minigap $\Delta$ in the junction area, which is the level spacing between the lowest-energy electron-like and hole-like ABSs, shall oscillate with $\varphi$ \cite{Beenakker_full_trans}:
\begin{equation}
\Delta=\Delta_0\sqrt{1-D\sin^2(\varphi/2)}
\end{equation}
where $\Delta_0$ is the maximal value of the minigap, $D$ is the total transmission coefficient of the junction, including those of the two S-N interfaces and that of the N part.

The oscillation of $\Delta$ with varying $\varphi$ is depicted in Fig. 4. It can be seen that the more transparent the junction is, the more oscillatory the minigap will be. But it is very difficult to reach full gap-closing --- the minigap remains significantly open even if the total transmission coefficient $D$ is as high as 0.99 (the red curves).

\begin{figure}
\includegraphics[width=0.95 \linewidth]{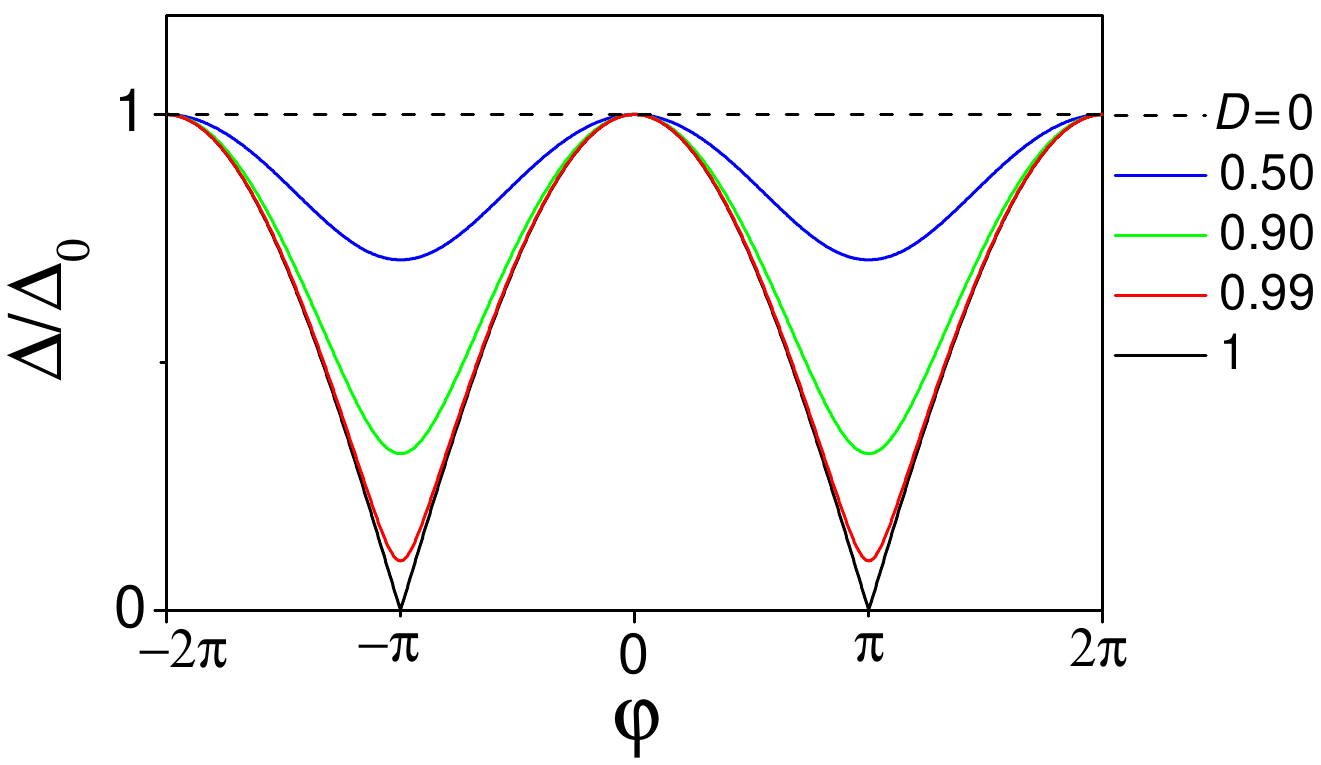}
\caption{\label{fig:fig4} {(color online) The oscillations of minigap $\Delta$ with varying phase difference $\varphi$ in S-N-S type Josephson junctions with different total transmission coefficient $D$. }}
\end{figure}

The local phase difference $\varphi (\phi, x)$ in Eq. 1 is controlled by the magnetic flux in the junction area:
\begin{equation}
\varphi (\phi, x)=2\pi\frac{x\phi}{W\phi_0}\pm\pi\hspace {0.05 cm}{\rm int}(\frac{\phi}{\phi_0})
\end{equation}
where $\phi=BWH$ is the total magnetic flux in the junction of area $WH$, and $x$ is defined from $-W/2$ to $W/2$. The first term in Eq. 2 is the local phase difference caused by magnetic flux. The second term represents a $\pi$ phase jump whenever the total flux in junction crosses the nodes of the Fraunhofer pattern.

For the mechanism of $\pi$ phase jump, it is well known that for single Josephson junctions there exists a phase offset $\varphi_0$ which can be self-adjusted to minimize the total energy. As can be seen in the supplementary materials \cite{supplementary_materials}, in the flux range of the 0$^{\rm th}$ Fraunhofer peak, the lowest-energy state corresponds to the $\varphi_0=0$ state. But in the flux range of the 1$^{\rm st}$ Fraunhofer peak, the lowest-energy state corresponds to the $\varphi_0=\pm\pi$ state. A $\pi$ phase jump will thus take place at the nodes of the Fraunhofer pattern.

\subsection{Transparent or non-transparent? Dilemma and solution}

As illustrated in Fig. 4, the minigap $\Delta$ is less oscillatory with varying magnetic flux when $D\rightarrow 0$ (the non-transparent limit), but becomes most oscillatory when $D\rightarrow 1$ (the transparent limit). The observation of significant $dV/dI_{\rm b}$ oscillation rules out that our Josephson junction is in the non-transparent limit --- a conclusion which is in sharp contrast to the one drawn from the Fraunhofer pattern in Fig. 2. To solve this dilemma, we have to believe that there exist two distinct CPRs, corresponding to two different supercurrents as illustrated in Fig. 1(b). One CPR corresponds to the majority supercurrent flowing through the bulk of Bi$_2$Te$_3$ in the non-transparent limit \cite{Qu_SR}, resulting in the measured Fraunhofer pattern of the non-transparent type. The other CPR corresponds to the supercurrent flowing through highly transparent channels, presumably the ABSs on the surface of Bi$_2$Te$_3$. The contact resistance depends sensitively on the surface electron density of states, hence a highly oscillatory $dV/dI_{\rm b}$ was detected.

In the transparent limit $D=1$, Eq. 1 becomes:
\begin{equation}
\Delta=\Delta_0|\cos (\varphi/2)|
\end{equation}
In this limit, the minigap undergoes complete closing at the Fraunhofer nodes.

In the next sections, we will analyze the data of the contact resistance quantitatively and qualitatively, to convince the readers that the minigap undergoes completely closing and that our Josephson junctions are indeed fully transparent.

\subsection{Simulation of the $dV/dI_{\rm b}-I_{\rm b}$ curves by using the BTK theory}

Although the significant oscillation of the contact resistance infers that the Pb-Bi$_2$Te$_3$-Pb Josephson junctions are in the transparent limit, the high value of the contact resistance itself tells that the transport process across the Pd-Bi$_2$Te$_3$ interface is rather non-transparent. We note that there was no intentionally made barrier at this interface. The remnant photoresist on the Bi$_2$Te$_3$ surface, if present, is $\sim$1 nm thick or less as revealed by atomic force microscopy studies. Such an interface should allow the happening of direct tunneling of single quasiparticles as well as the two-particle Andreev reflection process. These processes are usually described by the Blonder-Tinkham-Klapwijk (BTK) theory \cite{BTK}.

It is not quite straightforward that the BTK theory can be applied to describe the single-particle and two-particle processes across the Pd-Bi$_2$Te$_3$ interface here, since the superconducting gap in Bi$_2$Te$_3$ is not a standard one but an induced minigap between the lowest-energy electron-like and hole-like Andreev bound states (or continuums in the dirty regime). Nevertheless, while the electron-like and hole-like ABSs (or continuums) at the Bi$_2$Te$_3$ surface mediate the Josephson supercurrents between the two Pb electrodes, the same ABSs (or continuums) also define an energy window within which the two-particle process can occur between the Pd electrode and the superconducting bath (e.g., firstly to the superconducting bulk of Bi$_2$Te$_3$ \cite{Qu_SR}, then eventually to the Pb electrodes). Although a rigours theory is still needed to treat this special case, our numerical simulation below demonstrates that the BTK formalism works well to describe the single-particle and two-particle processes across the Pd-Bi$_2$Te$_3$ interface.

For a single conduction channel connecting a normal metal reservoir N and a superconductor reservoir S, the current can be expressed as:
\begin{equation}
I=\frac{G_{\rm Q}}{e}\int dE[1-b(E)+a(E)] [f(E)-f(E-eV)]
\end{equation}
where $G_{\rm Q}=2e^2/h$ is the quantum conductance, $b=r_{ee}^2$ is the normal (electron-to-electron) reflection coefficient to suppress the current, $a=r_{eh}^2$ is the Andreev (electron-to-hole) reflection coefficient to increase the current, and $f(E)=1/[1+\exp(E/k_{\rm B}T)]$ is the Fermi distribution function of the left/right reservoirs.

According to the BTK theory:
\begin{equation}
\begin{aligned}
&1-b\left(E\right)+a\left(E\right) \\
&=\left\{
\begin{aligned}
\frac{2\Delta^2}{{\left(eV\right)}^2+{\left(1+2Z_{\rm Pd-TI}^2\right)}^2\left(\Delta^2-{\left(eV\right)}^2\right)} & &(eV<\Delta)\\
\frac{2eV}{eV+{\left(1+2Z_{\rm Pd-TI}^2\right)}^2\sqrt{\left({\left(eV\right)}^2-\Delta^2\right)}} & &(eV>\Delta)
\end{aligned}
\right.
\end{aligned}
\end{equation}
where $\Delta$ is the minigap determined by the total transmission coefficient $D$ of the Pb-Bi$_2$Te$_3$-Pb Josephson junction via Eq. 1, and $Z_{\rm Pd-TI}$ is the barrier strength of the Pd-Bi$_2$Te$_3$ interface,  related to the transmission coefficient $D_{\rm Pd-TI}$ of the interface via $D_{\rm Pd-TI}=1/(1+Z_{\rm Pd-TI}^2)$.

\begin{figure}
\includegraphics[width=1 \linewidth]{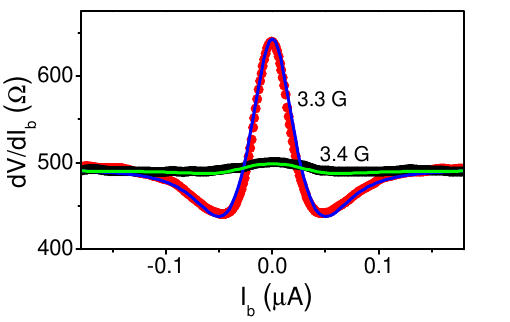}
\caption{\label{fig:fig5} {(color online) Simulation of the $dV/dI_{\rm b}-I_{\rm b}$ data measured by the central Pd electrode of device \#2. The red and black symbols are the same data as shown in Fig. 3(o), and the blue and green lines are the simulations by assuming the Pd electrode is a geometric point. The parameters used in the simulations are: $T$=50 mK, $Z_{\rm Pd-TI}$=3.3, $N$=670, $\Delta_{\rm 3.3 G}=9.25 \mu$V, and $\Delta_{\rm 3.4 G}=1.8 \mu$V. }}
\end{figure}

In the real case, there exist multiple channels at the Pd-Bi$_2$Te$_3$ interface. For simplicity, we assume that the transmission coefficient of each channel is the same. The total current is then:
\begin{equation}
I=\frac{G_{\rm Q}}{e}N\int dE[1-b(E)+a(E)][f(E)-f(E-eV)]
\end{equation}
where $N$ is the number of parallel channels.

With the above formula, we can simulate the $I-V$ curve, the $dV/dI_{\rm b}-V_{\rm b}$ curve, as well as the $dV/dI_{\rm b}-I_{\rm b}$ curve of the Pd-Bi$_2$Te$_3$ interface. Figure 5 shows the simulations on the $dV/dI_{\rm b}-I_{\rm b}$ data of device \#2 shown in Fig. 3(o). The fitting parameters are $Z_{\rm Pd-TI}=3.3$, $N=670$ and $T=50$ mK. The minigap is 9.25 $\mu$V before closing at $B=3.3$ G (the blue line), and 1.8 $\mu$V after closing at $B=3.4$ G (the green line). The large height reduction of the gap structure yields a very high transmission coefficient of $D=1-(\Delta/\Delta_0)^2\approx 1-(\Delta_{3.4 \rm G}/\Delta_{3.3 \rm G})^2=0.96$ for the Pb-Bi$_2$Te$_3$-Pb Josephson junction. This value of $D$ is comparable to that of a single-atomic-layer Josephson junction \cite{G. H. Lee} and an atomic superconducting point contact \cite{Rocca}. It is in sharp contrast to the transmission coefficient of the Pd-Bi$_2$Te$_3$ interface of the central Pd electrode in device \#2, where $Z_{\rm Pd-TI}=3.3$ and hence the transmission coefficient is only $D_{\rm Pd-TI}=1/(1+Z_{\rm Pd-TI}^2)=0.084$.

\subsection{The finite-size effect of the Pd electrodes}

In the above simulation we assume that the Pd electrode is a geometric point. In the real case, however, the Pd electrode always picks up signals within an area where the minigap remains unclosed at most places, such that the measured signature of gap closing is largely diminished. In other words, the true minigap at the center of the junction should be much smaller than 1.8 $\mu$V when the magnetic flux reaches the Fraunhofer nodes. We therefore believe that the total transmission coefficient $D$ of our S-TI-S junction is much higher than 0.96, being already in the fully transparent limit.

\begin{figure}
\includegraphics[width=0.80 \linewidth]{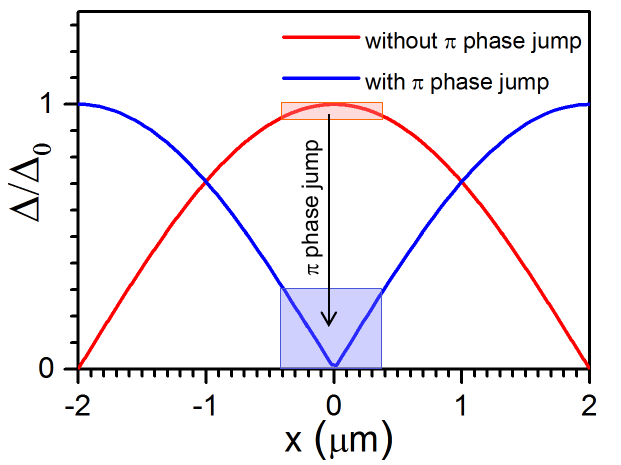}
\caption{\label{fig:fig6} {(color online) The position dependence of mingap in a fully transparent junction at magnetic flux near the first Fraunhofer node. Red curve: before $\pi$ phase jump, i.e., $\phi=\phi_0^-$. Blue curve: after $\pi$ phase jump, i.e., $\phi=\phi_0^+$. The width of the light-red and light-blue rectangles represents the diameter of the central Pd electrode along the $x$ direction, and the heights of the rectangles indicate the ranges of minigap that the Pd electrode of given width probes. }}
\end{figure}

In the following, we demonstrate that the remaining gap structure on the black/green curves in Fig. 5 can be well ascribed to the finite-size effect of the Pd electrode while complete gap closing already occurs at $x=0$. From Eqs. 2 and 3, the position dependence of the minigap in the full transparent limit can be calculated. The results before and after $\pi$ phase jump in the vicinity of the first Fraunhofer node are plotted in Fig. 6. It can be seen that strict gap closing takes place only at the ends of the junction before the $\pi$ phase jump, and at the center after the $\pi$ phase jump. Obviously, the central Pd electrode of device \#2 with width $\sim$800 nm illustrated by the light blue rectangle in Fig. 6 picks up not only the gap-closing signal at the center of the junction, but also the gap-opening signal in the vicinity area. Therefore, the measured signature of gap closing in real experiment is diminished by the finite size effect.

\begin{figure}
\includegraphics[width=0.85 \linewidth]{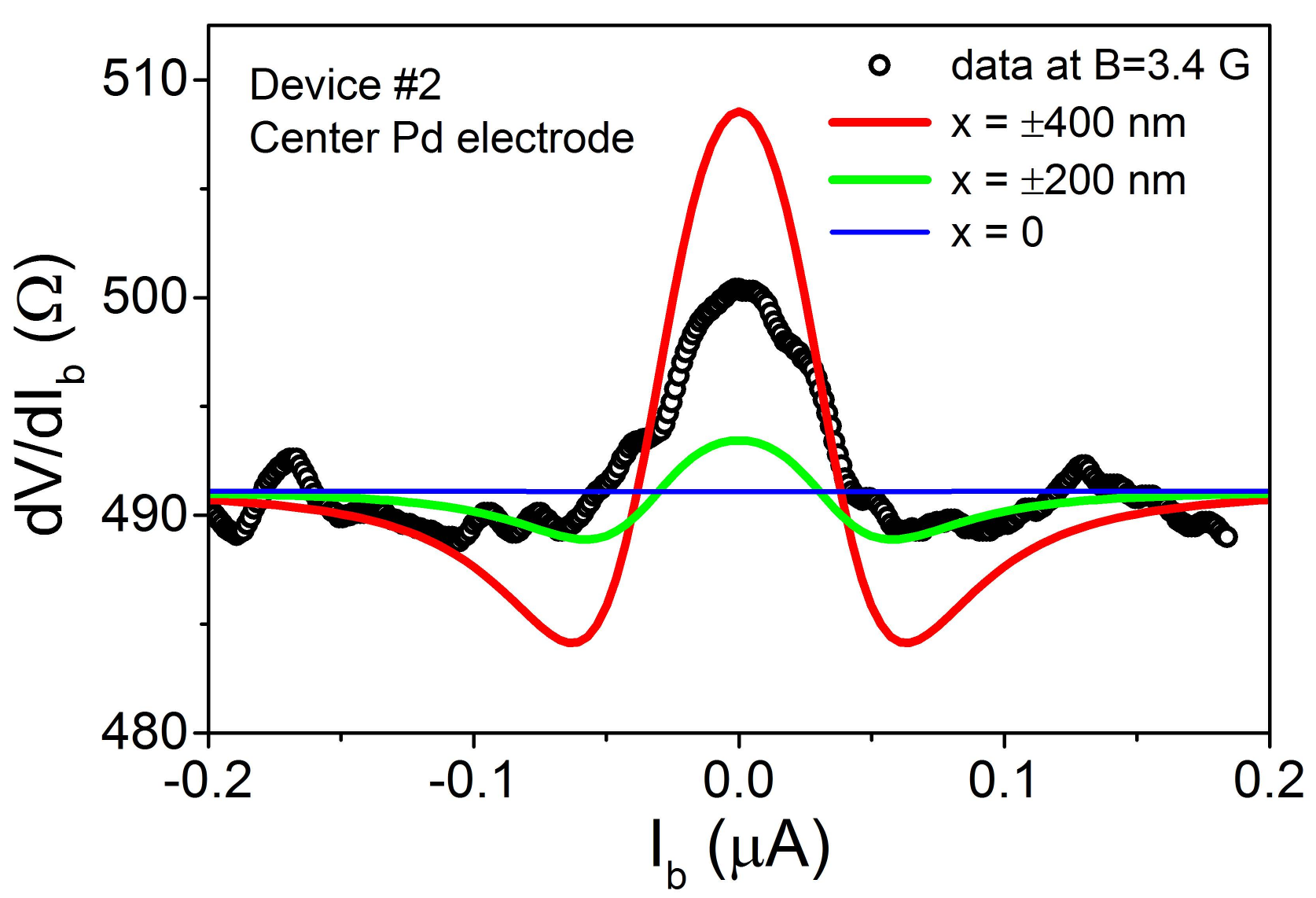}
\caption{\label{fig:fig7} {(color online) Comparison between the remaining gap structure measured by the central Pd electrode of diameter $d=800$ nm on device \#2 at $B=3.4$ G and the calculated ones at positions $x=0$, $\pm$200 nm and $\pm$400 nm, assuming the Pd-Bi$_2$Te$_3$-Pb junction is in the fully transparent limit. The other parameters used in the calculation are the same as in Fig. 5: $T$=50 mK, $Z_{\rm Pd-TI}$=3.3, $N$=670, and $\Delta_0\approx\Delta_{3.3 \rm G}=9.25 \mu$V. }}
\end{figure}

In Fig. 7 we show the $dV/dI_{\rm b}-I_{\rm b}$ curves at $x=0, \pm$200 nm and $\pm$400 nm expected in the fully transparent limit, in comparison with the experimental data at $B=3.4$ G taken from the central Pd electrode of device \#2 whose diameter is $\sim$800 nm. The results unambiguously show that the remaining tiny feature of gap in the experimental data can be a finite-size effect of the Pd electrode.

\subsection{Simulations of the magnetic flux dependencies of zero-bias $dV/dI_{\rm b}$ in the fully transparent limit and after taking into account the finite-size effect of the Pd electrodes}

Taking the finite-size effect of the Pd electrodes into account, we can further simulate the magnetic flux dependence of the data in the fully transparent limit $D=1$.

\begin{figure*}
\includegraphics[width=0.8 \linewidth]{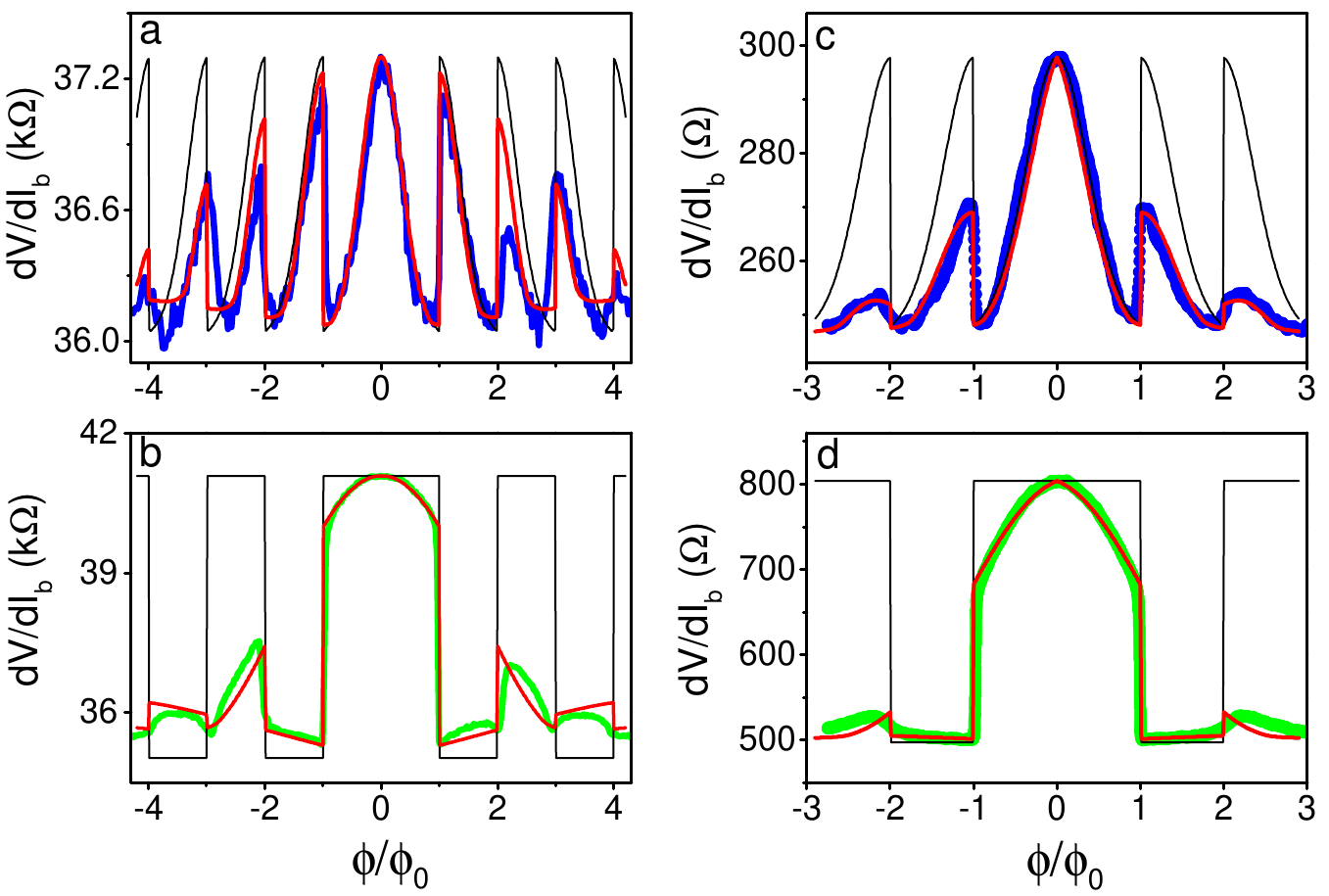}
\caption{\label{fig:fig8} {(color online) Simulations of the magnetic flux dependence of $dV/dI_{\rm b}$ in the fully transparent limit. The black lines are the simulations without considering the size effect of the Pd electrodes. The red lines are the simulations after considering the size effect.
(\textbf{a}) Simulating the data in Fig. 3(e) with fitting parameters $T$=50 mK, $Z_{\rm Pd-TI}$=6.5, $N$=21, $\Delta_0=12 \mu$V. The diameter of the side Pd electrode in device \#1 is $d=$600 nm.
(\textbf{b}) Simulating the data in Fig. 3(f) with fitting parameters $T$=50 mK, $Z_{\rm Pd-TI}$=6.9, $N$=19, $\Delta_0=10.3 \mu$V. The diameter of the central Pd electrode in device \#1 is $d=$660 nm.
(\textbf{c}) Simulating the data in Fig. 3(l) with fitting parameters $T$=50 mK, $Z_{\rm Pd-TI}$=2.1, $N$=550, $\Delta_0=12 \mu$V. The diameter of the side Pd electrode in device \#2 is $d=$770 nm.
(\textbf{d}) Simulating the data in Fig. 3(m) with fitting parameters $T$=50 mK, $Z_{\rm Pd-TI}$=3.3, $N$=670, $\Delta_0=12 \mu$V. The diameter of the central Pd electrode in device \#2 is $d=$800 nm.}}
\end{figure*}

Because the superconducting gap underneath the Pd electrode varies specially, the normal-reflection coefficient $b$ and the Andreev-reflection coefficient $a$ are functions of position $x$. The total current can be rewritten as:
\begin{equation}
I=\frac{G_{\rm Q}}{e}\int n(x)dx\int dE[1-b(E)+a(E)][f(E)-f(E-eV)]
\end{equation}
where $n(x)$ is the number of channels per unit area (i.e., channel density). For simplicity, we assume the channel density is a constant $n_0$. The formula becomes:
\begin{equation}
I=\frac{G_{\rm Q}}{e}n_0\int dx\int dE[1-b(E)+a(E)][f(E)-f(E-eV)]
\end{equation}

With Eqs. 5 and 8, the magnetic field dependence of $dV/dI_{\rm b}$ can be simulated. Plotted in Figs. 8(a), (b), (c) and (d) are the simulations to the experimental data shown in Figs. 3(e), (f), (l) and (m), respectively. The blue and green dots are the experimental data. The black lines are the simulations assuming zero-sized electrodes located strictly at $x=\pm W/2$ (the ends) or $x=0$ (the center). The red lines in Figs. 8(a) and (b) are the simulations for electrodes of diameter 600 nm at the end and 660 nm at the center of device \#1, respectively. And the red lines in Figs. 8(c) and (d) are the simulations for electrodes of diameter $\sim$770 nm at the end and $\sim$800 nm at the center of device \#2, respectively. The excellent agreement between the data and the simulations supports the conclusion that the Pb-Bi$_2$Te$_3$-Pb junctions are in the fully transparent limit, and that complete gap closing indeed occurs.

\subsection{Further discussions}

The total transmission coefficient $D$ of an Pb-TI-Pb Josephson junction relies on the transmission coefficients of the two Pb-TI interfaces and that of the TI part. Although in this experiment we did not measure the barrier strength $Z_{\rm TI-Pb}$ of single Bi$_2$Te$_3$-Pb interfaces, our early study on Bi$_2$Se$_3$-Sn interfaces prepared via the same techniques shows that the barrier strength is not close to 0 at all, but remaining as high as 0.6 even after additional reactive ion etching was applied to reduce the remnant photoresist \cite{ZBCP Fan Yang}. If taking $Z_{\rm TI-Pb}=0.6$, single Bi$_2$Te$_3$-Pb interface yields a transmission coefficient of $\sim 1/(0.6^2+1)=0.74$, being already significantly smaller than 1. In addition to the interface scattering, the Bi$_2$Te$_3$ part in our junction is 1 $\mu$m in length. We therefore conclude that full transparency of quasiparticle transport in our Pb-Bi$_2$Te$_3$-Pb junctions could hardly be observed if without a mechanism to prohibit the happening of backscatterings. It is due to the single-helicity nature of the surface states that the backscatterings are completely suppressed and perfect Andreev reflections are enabled between the two Bi$_2$Te$_3$-Pb interfaces, leading to the formation of MZMs when $\pi$ phase difference is reached. In contrast, the charge transport along the perpendicular direction across the Pd-Bi$_2$Te$_3$ interface has nothing to do with the zero mode formation, and thus is not protected to be fully transparent. In fact, the barrier strength $Z_{\rm Pd-TI}$ of single Pd-Bi$_2$Te$_3$ interface ranges from 2.1 to 6.9 in this experiment (see the fitting parameters in Fig. 8). Taking the central Pd electrode of device \#2 as an example, whose $Z_{\rm Pd-TI}=3.3$, its transmission coefficient is only $1/(Z_{\rm Pd-TI}^2+1)=0.084$.

\section{SUMMARY}

We have performed contact resistance measurement on single Josephson junctions constructed on Bi$_2$Te$_3$ surface. Complete gap closing is inferred from the data, indicating the existence of topologically protected full transparency of charge transport through the surface states of Bi$_2$Te$_3$. Our results support the Fu-Kane proposal that Majorana zero modes can be hosted in Josephson junctions constructed on the 2D surface of 3D TIs. Based on such a 2D platform, universal topological quantum computation could be further implemented \cite{FuPRX}.

\vspace {0.5 cm}

{\it Note added}: The first version of this paper was posted on arXiv (arXiv:1603.04540v1). During revision we noticed that a similar work was recently carried out by Sch\"{u}ffelgen and coworkers \cite{boosting}, in which the authors show that for a Nb-Bi$_2$Te$_3$-Nb Josephson junction containing a few tens nm of Bi$_2$Te$_3$ in length, the total transmission coefficient is as high as 0.95.

\vspace {0.5 cm}

%\begin{acknowledgments}
\noindent\textbf{Acknowledgments} We would like to thank L. Fu, G. M. Zhang, T. Xiang, X. C. Xie, Q. F. Sun, R. Du, Z. Fang, X. Dai, X. Hu and Z. G. Cheng for fruitful discussions. This work was supported by the National Basic Research Program of China from the MOST grants 2016YFA0300601, 2017YFA0304700, 2015CB921402, 2011CB921702 and 2009CB929101, by NSFC grants 11527806, 91221203, 11174340, 11174357, 91421303, 11774405 and by the Strategic Priority Research Program B of the Chinese Academy of Sciences grant No. XDB07010100.
%\end{acknowledgments}

\begin{widetext}


\section*{Supplementary material}

# Supplementary Materials for
# "Protected gap closing in Josephson junctions constructed on Bi$_2$Te$_3$ surface"

Zhaozheng Lyu[1,*], Yuan Pang[1,*], Junhua Wang[1,*], Guang Yang[1], Jie Fan[1], Guangtong Liu[1], Zhongqing Ji[1], Xiunian Jing[1,2], Changli Yang[1,2], Fanming Qu[1,3] and Li Lu[1,2]†

[1] Beijing National Laboratory for Condensed Matter Physics,
Institute of Physics, Chinese Academy of Sciences; School of Physical Sciences,
University of Chinese Academy of Sciences, Beijing 100190, People's Republic of China

[2] Collaborative Innovation Center of Quantum Matter, Beijing 100871, People's Republic of China

[3] CAS Center for Excellence in Topological Quantum Computation, University of Chinese Academy of Sciences, Beijing 100190, People's Republic of China

* These authors contributed equally to this work.

† Corresponding authors: lilu@iphy.ac.cn

Contents:
1. Estimation of effective junction area after considering flux penetration and compression
2. The mechanism of $\pi$ phase shift in single Josephson junctions
3. The expected Fraunhofer patterns of supercurrent in non-transparent and transparent Josephson junctions

# 1. Estimation of the effective junction area after considering flux penetration and compression

In Fig. S1 we illustrate how the effective junction area of the device shown in Fig. 1 of the main manuscript is calculated after considering flux penetration and compression as well as the stray distribution of the supercurrent.

As illustrated in the upper left panel of Fig. S1, below $H_{c1}$ the magnetic field cannot penetrate into the bulk of superconducting Pb film other than within the penetration depth. The flux lines will therefore be pushed away from the Pb electrodes and focused in the junction area, making the effective junction area larger as defined by the white dashed lines, being slightly less than the area defined by the red dashed line.

On the other hand, we know that superconductivity will be induced on the TI surface by proximity effect, spreading away to a distance of micron [Qu, F. et al., Sci. Rep. 2, 339 (2012); Yang, F. et al., Phys. Rev. B 86, 134504 (2012)]. Therefore, the TI surface near Pb electrodes will be able to carry stray supercurrents.

After taking into consideration of the above facts, we employed an empirical method of estimating the effective junction area as shown in the right panel of Fig. S1. We believe that the effective junction area is close to the one enclosed by the blue dashed line, which is approximately equal to the area defined by the red dashed line. By comparing with the effective area drawn from the measured Fraunhofer patterns, the effective area estimated in way turns out to be accurate, with an error less than 20%.

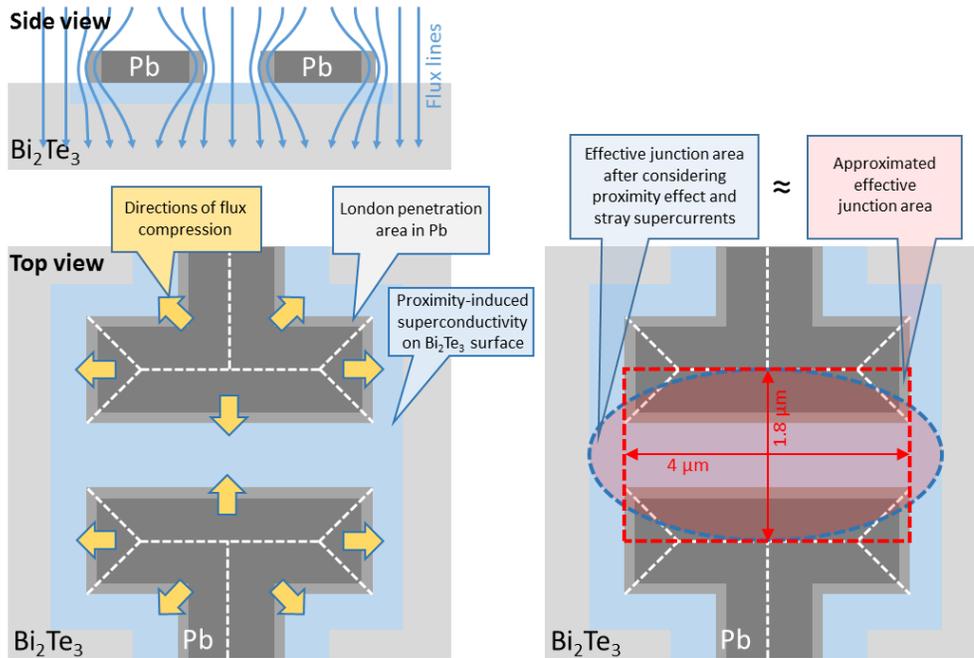

**FIG. S1 | Upper left panel:** Side view for flux compression. **Lower left panel:** Top view for flux compression. The flux in each area defined by white dashed lines will be compressed away along the arrowed directions. **Right panel:** Our empirical method of estimating the effective junction area after considering flux compression and stray supercurrent distribution.

## 2. The mechanism of π phase shift in a single Josephson junction

In a single Josephson junction, the phase difference $\varphi$ across the junction can be freely adjusted according to the criteria of minimizing the total energy of the junction. In the following we show that a π phase jump will automatically occur at the node fields of the Fraunhofer pattern.

In the non-transparent limit $D \to 0$ (where $D$ is the transmission coefficient), the current-phase relation of the junction is sinusoidal, such that the total free energy of the junction can be expressed as:

$$E_J = \begin{cases} \dfrac{\Delta_0}{W} \int_{-\frac{W}{2}}^{\frac{W}{2}} \left[1 - \cos(\dfrac{2\pi\phi}{\phi_0})\right] dx & \text{(without adjustable phase offset)} \\ \dfrac{\Delta_0}{W} \int_{-\frac{W}{2}}^{\frac{W}{2}} \left[1 - \cos(\dfrac{2\pi\phi}{\phi_0} \pm \pi a)\right] dx & \text{(with an adjustable phase offset)} \end{cases}$$

where $\Delta_0$ is the maximum value of gap, $\phi = BHx$ is the magnetic flux in the junction area, $B$ the magnetic field, $H$ the effective length, $W$ the width of the junction, $\phi_0$ the flux quantum, and $\varphi_0 = \pi a$ is a phase offset which can be freely adjusted according to the criteria of minimizing the total energy.

The above equations can be rewritten as:

$$E_J = \begin{cases} \dfrac{\Delta_0}{2} \int_{-1}^{1} \left[1 - \cos(\dfrac{\pi\phi'}{\phi_0} x')\right] dx' & \text{(without adjustable phase offset)} \\ \dfrac{\Delta_0}{2} \int_{-1}^{1} \left[1 - \cos(\dfrac{\pi\phi'}{\phi_0} x' \pm \pi a)\right] dx' & \text{(with an adjustable phase offset)} \end{cases}$$

where $x' = 2x/W$ and $\phi' = BHW$.

Shown in Fig. S2 (a) is the calculated $E_J$ at different values of $a$. It can be seen that, in order for the system to trace the lowest energy state, a π phase shift ($a$ jumping by 1) will happen whenever the applied flux crosses the Fraunhofer nodes.

In the transparent limit $D \to 1$ whose energy-phase relations and current-phase relation are shown in Fig. S3 (a), including the trivial ballistic case and non-trivial 4π-period case but suffered with quasiparticle poisoning, the energies of the Andreev bound states take the form of $|\Delta_0 \cos(\varphi/2)|$, so that the total free energy of the junction can be expressed as:

$$E_J = \begin{cases} \dfrac{\Delta_0}{W} \int_{-\frac{W}{2}}^{\frac{W}{2}} \left[1 - \left|\cos(\dfrac{\pi\phi}{\phi_0})\right|\right] dx & \text{(without adjustable phase offset)} \\ \dfrac{\Delta_0}{W} \int_{-\frac{W}{2}}^{\frac{W}{2}} \left[1 - \left|\cos(\dfrac{\pi\phi}{\phi_0} \pm \dfrac{\pi}{2} a)\right|\right] dx & \text{(with an adjustable phase offset)} \end{cases}$$

or

$$E_J = \begin{cases} \dfrac{\Delta_0}{2} \int_{-1}^{1} \left[1 - \left|\cos(\dfrac{\pi\phi'}{2\phi_0} x')\right|\right] dx & \text{(without adjustable phase offset)} \\ \dfrac{\Delta_0}{2} \int_{-1}^{1} \left[1 - \left|\cos(\dfrac{\pi\phi'}{2\phi_0} x' \pm \dfrac{\pi}{2} a)\right|\right] dx' & \text{(with an adjustable phase offset)} \end{cases}$$

As shown in Fig. S2 (b), again, in order for the system to trace the lowest energy state, a phase shift of π/2 (i.e., $a$ jumping by 1) will happen whenever the applied flux crosses the

Fraunhofer nodes.

To summarize, in both the non-transparent and transparent limits, the phase offset of a single Josephson junction will undergo a jumping (with a proper phase shift, e.g., $\pi$ or $\pi/2$) when the magnetic flux crosses the Fraunhofer nodes.

The phase jumping associates with a redistribution of the Josephson supercurrent in the junction area, as well as a redistribution of flux of amount $\phi_0$ in the junction area.

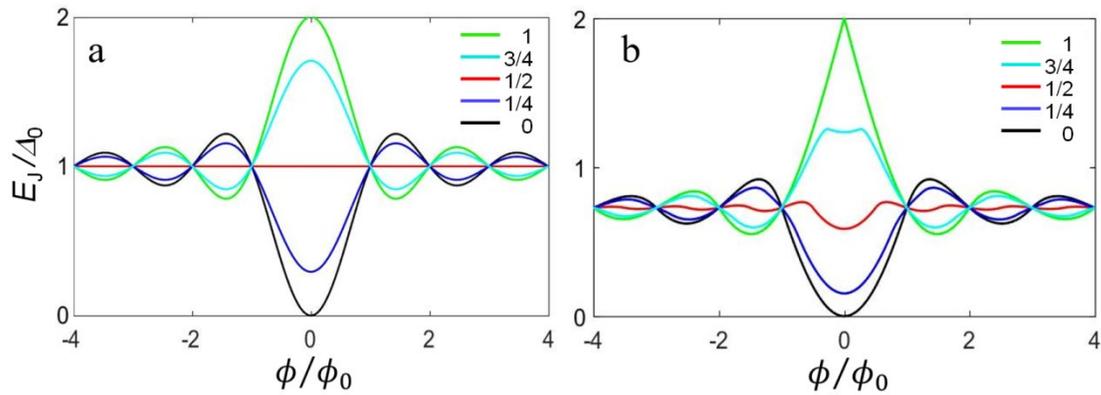

**FIG. S2 |** Total free energy of a single Josephson junction as a function of magnetic flux at different values of adjusting parameter *a*, in the non-transparent limit $D=0$ **(a)**; and the transparent limit $D=1$ **(b)**.

## 3. The expected Fraunhofer patterns of supercurrent in non-transparent and transparent Josephson junctions

As shown in the main manuscript and in the previous section, the phase difference across the junction is:

$$\varphi(x) = 2\pi \frac{BHx}{\phi_0} \pm \pi \operatorname{int}\left(\frac{\phi}{\phi_0}\right) + \varphi_0$$

where $-W/2 < x < W/2$, $\phi = BHW$, $B$ is the applied magnetic field, $H$ the effective length, $W$ the width of the junction, and $\varphi_0$ is a phase offset which can be freely adjusted according to the criteria of minimizing the total energy of the junction.

In the non-transparent limit $D=0$, the supercurrent density at position $x$ is:

$$j_{s1} = j_c \sin\varphi(x)$$

The total supercurrent of the junction is the integral of $j_{s1}$ along the x direction:

$$I_1 = \int_{-W/2}^{W/2} j_{s1}\, dx = \int_{-W/2}^{W/2} j_c \sin[\varphi(x)]\, dx = \pm I_c \sin(\varphi_0) \frac{\sin(\pi\frac{\phi}{\phi_0})}{\pi\frac{\phi}{\phi_0}}$$

As mentioned earlier, $\varphi_0$ is adjustable and should be determined according to the criteria of minimizing the total energy of the junction. It results in the $\pi$ phase jump at the Fraunhofer nodes, by which the critical supercurrent keeps to be positive

$$I_1 = I_c \left|\frac{\sin(\pi\frac{\phi}{\phi_0})}{\pi\frac{\phi}{\phi_0}}\right|$$

This is the well-known Fraunhofer diffraction pattern of critical supercurrent for non-transparent Josephson junction.

In the transparent limit $D=1$, if the full transparency is protected by the Majoranas, both the energy-phase relation and the current phase relation will be $4\pi$–period. The Josephson supercurrent density is:

$$j_{s2} = j_c \sin[\varphi(x)/2]$$

And the Fraunhofer diffraction pattern becomes:

$$I_2 = I_c \left|\frac{\sin(\frac{\pi\phi}{2\phi_0})}{\frac{\pi\phi}{2\phi_0}}\right|$$

Theoretically, full transparency can also be reached in trivial ballistic Josephson junctions. In this case, the electron and hole-like Andreev bound states follow the functional forms of $E_\pm = \pm\Delta|\cos(\phi/2)|$ as represented by the dashed red line and the black line in Fig. S3(a), respectively. And the current-phase relation becomes fully skewed as represented by the blue line in Fig. S3(a), in the functional form of:

$$j_{s3} = j_c \sin\left[\frac{\varphi(x)}{2}\right] * \operatorname{sign}\{\cos[\varphi(x)/2]\}$$

where sign(x) is the function for getting the sign of variable x. Same functional form will be

followed if the 4π–period current-phase relation in the Majorana case is poisoned by quasiparticles.

With the above formula of critical supercurrent density, the Fraunhofer diffraction for trivial ballistic junction can be obtained:

$$I_3 = \int_{-W/2}^{W/2} j_c \sin\left[\frac{\varphi(x)}{2}\right] * \text{sign}\{\cos[\varphi(x)/2]\}\, dx$$

In Fig. S3 (b), the flux dependence of $I_1$, $I_2$ and $I_3$ are renormalized and plotted together with the measured Fraunhofer pattern of device #1 presented in the main manuscript.

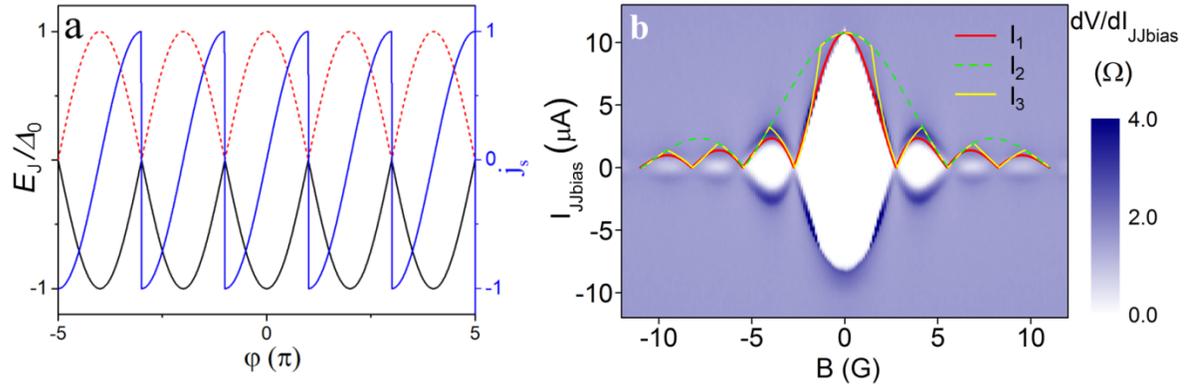

**FIG. S3 | (a)** The energy-phase relation of the electron-like (dashed red line) and hole-like (black line) Andreev bound states, together with the fully skewed current-phase relation (blue line) of a Josephson junction in the transparent limit D=1. **(b)** The expected Fraunhofer diffraction patterns of critical supercurrent in the non-transparent limit (red curve), the non-trivial 4π–period transparent limit (dashed green curve), and the trivial transparent limit or the non-trivial but quasiparticle poisoned state (yellow curve).


\end{widetext}


\begin{thebibliography}{10}

\bibitem{Majorana} E. Majorana, Nuovo Cimento {\bf 14}, 171-184 (1937).

\bibitem{Majorana return} F. Wilczek, Nature Phys. {\bf 5}, 614-618 (2009).

\bibitem{Search for Majorna} R. F. Service, Science {\bf 332}, 193-195 (2011).

\bibitem{Race for Majorana} M. Franz, Physics {\bf 3}, 24 (2010); M. Franz, arXiv: 1302.3641v1.

%ZBCPs
\bibitem{ZBCP Kouwenhoven} V. Mourik, K. Zuo, S. M. Frolov, S. R. Plissard, E. P. A. M. Bakkers, and L. P. Kouwenhoven, Science {\bf 336}, 1003-1007 (2012).

\bibitem{ZBCP H.Q.Xu} M. T. Deng, C. L. Yu, G. Y. Huang, M. Larsson, P. Caroff, and H. Q. Xu, Nano Lett. {\bf 12}, 6414-6419 (2012).

\bibitem{ZBCP shtrikman} A. Das, Y. Ronen, Y. Most, Y. Oreg, M. Herblum, and H. Shtrikman, Nature Phys. {\bf 8}, 887-895 (2012).

\bibitem{ZBCP Marcus} H. O. H. Churchill, V. Fatemi, K. Grove-Rasmussen, M. T. Deng, P. Caroff, H. Q. Xu, and C. M. Marcus, Phys. Rev. B {\bf 87}, 241401 (2013).

\bibitem{ZBCP van Harlingen} A. D. K. Finck, D. J. Van Harlingen, P. K. Mohseni, K. Jung, and X. Li, Phys. Rev. Lett. {\bf 110}, 126406 (2013).

\bibitem{ZBCP Yazdani} S. Nadj-Perge, I. K. Drozdov, J. Li, H. Chen, S. Jeon, J. Seo, A. H. MacDonald, B. A. Bernevig, and A. Yazdani, Science {\bf 346}, 602-607 (2014).

\bibitem{ZBCP Ando} S. Sasaki, M. Kriener, K. Segawa, K. Yada, Y. Tanaka, M. Sato, and Y. Ando, Phys. Rev. Lett. {\bf 107}, 217001 (2011).

\bibitem{ZBCP Marcus_2} M. T. Deng, S. Vaitiekenas, E. B. Hansen, J. Danon,  M. Leijnse, K. Flensberg, J. Nyg{\aa}rd, P. Krogstrup, and C. M. Marcus, Science {\bf 354}, 6319 (2016).

\bibitem{ZBCP Fan Yang} F. Yang, Y. Ding, F. Qu, J. Shen, J. Chen, Z. Wei, Z. Ji, G. Liu, J. Fan, C. Yang, T. Xiang, and L. Lu, Phys. Rev. B {\bf 85}, 104508 (2012).

%fractional Josephson effect
\bibitem{Rokhinson} L. P. Rokhinson, X. Liu, and J. K. Furdyna, Nature Phys. {\bf 8}, 795-799 (2012).

\bibitem{Wiedenmann} J. Wiedenmann, E. Bocquillon, R. S. Deacon, S. Hartinger, O. Herrmann, T. M. Klapwijk, L. Maier, C. Ames,  C. Br\"{u}ne, C. Gould, A. Oiwa, K. Ishibashi, S. Tarucha, H. Buhmann, and L. W. Molenkamp, Nature Commun. {\bf 7}, 10303 (2016).

%4pi CPR, exp
\bibitem{phase-sensitive10 and QP2 Yacoby} S. P. Lee, K. Michaeli, J. Alicea, and A. Yacoby, Phys. Rev. Lett. {\bf 113}, 197001 (2014).

\bibitem{Moler_PRL} I. Sochnikov, L. Maier, C. A. Watson, J. R. Kirtley, C. Gould, G. Tkachov, E. M. Hankiewicz, C. Br\"{u}ne, H. Buhmann, L. W. Molenkamp, and K. A. Moler, Phys. Rev. Lett. {\bf 114}, 066801 (2015).

\bibitem{phase-sensitive7 van Harlingen NatureComm} C. Kurter, A. D. K. Finck, Y. S. Hor, and D. J. Van Harlingen, Nature Commun. {\bf 6}, 7130 (2015).

\bibitem{Pang_CPB} Y. Pang, J. Wang, Z. Z. Lyu, G. Yang, J. Fan, G. T. Liu, Z. Q. Ji, X. N. Jing,
C. L. Yang, and Li Lu, Chin. Phys. B  {\bf 25}, 117402 (2016).

%4pi CPR, theory
\bibitem{4pi1 A Yu Kitaev} A. Yu Kitaev, Phys.-Usp. {\bf 44}, 131 (2001).

\bibitem{4pi2 LowTempPhy} H. J. Kwon, V. M. Yakovenko, and K. Sengupta, Low Temp. Phys. {\bf 30}, 613 (2004).

\bibitem{4pi3 and rf1 Fu Liang 2009 PRB} L. Fu and C. L. Kane, Phys. Rev. B {\bf 79}, 161408 (2009).

\bibitem{4pi4 and rf2 nanowire} R. M. Lutchyn, J. D. Sau, and S. Das Sarma, Phys. Rev. Lett. {\bf 105}, 077001 (2010).

\bibitem{4pi5 and rf3 Beenakker} M. Diez, I. C. Fulga, D. I. Pikulin, M. Wimmer, A. R. Akhmerov, and C. W. J. Beenakker, Phys. Rev. B {\bf 87}, 125406 (2013).

\bibitem{4pi6 and rf4 Kane} B. J. Wieder, F. Zhang, and C. L. Kane, Phys. Rev. B {\bf 89}, 075106 (2014).

\bibitem{current reverse and QP1} C. W. J. Beenakker, D. I. Pikulin, T. Hyart, H. Schomerus, and J. P. Dahlhaus, Phys. Rev. Lett. {\bf 110}, 017003 (2013).

%MBS, gap closing theory1
\bibitem{Fu 2008 PRL} L. Fu and C. L. Kane, Phys. Rev. Lett. {\bf 100}, 096407 (2008).

\bibitem{Fu 2013 PRB} A. C. Potter and L. Fu, Phys. Rev. B {\bf 88}, 121109 (2013).

\bibitem{ZBCP Hao Zhang} H. Zhang, C. -X. Liu, S. Gazibegovic, D. Xu, J. A. Logan, G. Wang, N. van Loo, J. D. S. Bommer, M. W. A. de Moor, D. Car, R. L. M. O. het Veld, P. J. van Veldhoven, S. Koelling, M. A. Verheijen, M. Pendharkar, D. J. Pennachio, B. Shojaei, J. S. Lee, C. J. Palmstr{\o}m, E. P. A. M. Bakkers, S. Das Sarma, L. P. Kouwenhoven, Nature doi:10.1038/nature26142.

\bibitem{supplementary_materials} See Supplemental Material at [URL will be inserted by publisher] for more discussions.

\bibitem{Beenakker_full_trans} C. W. J. Beenakker, {\it Transport phenomena in Mesoscopic Systems}, H. Fukuyama and T. Ando, eds. Springer, Berlin (1992).

\bibitem{Qu_SR} F. Qu, F. Yang, J. Shen, Y. Ding, J. Chen, Z. Q. Ji, G. T. Liu, J. Fan, X. N. Jing, C. L. Yang and L. Lu, Sci. Rep. {\bf 2}, 339 (2012).

\bibitem{BTK} G. E. Blonder, M. Tinkham, and T. M. Klapwijk, Phys. Rev. B {\bf 25 }, 4515 (1982).

\bibitem{G. H. Lee} G. H. Lee, S. Kim, S. H. Jhi, and H. J. Lee, Nature Commun. {\bf 6}, 6181 (2015).

\bibitem{Rocca} M. L. Della Rocca, M. Chauvin, B. Huard, H. Pothier, D. Esteve, and C. Urbina, Phys. Rev. Lett. {\bf 99}, 127005 (2007).

\bibitem{FuPRX} S. Vijay, T. H. Hsieh, and L. Fu, Phys. Rev. X {\bf 5}, 041038 (2015).

\bibitem{boosting} P. Sch\"{u}ffelgen, Da. Rosenbach, C. Li, T. Schmitt, M. Schleenvoigt, A. R. Jalil, J. K\"{o}lzer, M. Wang, B. Bennemann, Um. Parlak, L. Kibkalo, M. Luysberg, G. Mussler, A. A. Golubov, A. Brinkman, T. Sch\"{a}pers, D. Gr\"{u}tzmacher, arXiv:1711.01665.

\end{thebibliography}
\end{document}